\documentclass[utf8]{frontiersFPHY} 

\setcitestyle{square} 
\usepackage{url,hyperref,lineno,microtype,subcaption}
\hypersetup{colorlinks=true, linkcolor=magenta, urlcolor=blue}
\usepackage{xcolor}

\usepackage[onehalfspacing]{setspace}
\usepackage{xcolor}

\def\keyFont{\fontsize{8}{11}\helveticabold }
\def\firstAuthorLast{Argall {et al.}} 
\def\Authors{Matthew R. Argall\,$^{1,*}$, Colin R. Small\,$^{2}$, Samantha Piatt\,$^{2}$, Liam Breen\,$^{2}$, Marek Petrik\,$^{2}$, Kim Kokkonen\,$^{3}$, Julie Barnum\,$^{3}$, Kristopher Larsen\,$^{3}$, Frederick D. Wilder\,$^{3}$, Mitsuo Oka\,$^{4}$, William R. Paterson\,$^{5}$, Roy B. Torbert\,$^{1,6}$, Robert E. Ergun\,$^{3}$, Tai Phan\,$^{4}$, Barbara L. Giles\,$^{5}$, and James L. Burch\,$^{7}$}
\def\Address{$^{1}$Space Science Center, EOS, University of New Hampshire, 8 College Rd., Durham, USA \\
$^{2}$Department of Computer Science, University of New Hampshire, Durham, NH, USA \\
$^{3}$Laboratory for Atmospheric and Space Physics, University of Colorado at Boulder, Boulder, CO, USA \\
$^{4}$University of California at Berkeley, Berkeley, CA, USA \\
$^{5}$Goddard Space Flight Center, NASA, Greenbelt, MD, USA \\
$^{6}$EOS-SwRI, Southwest Research Institute, Durham, NH, USA \\
$^{7}$Southwest Research Institute, San Antonio, TX, USA }
\def\corrAuthor{Matthew R. Argall, Institute for the Study of Earth, Oceans, and Space, Space Science Center, University of New Hamsphire, 8 College Rd., Durham, NH, 03824, USA}
\def\corrEmail{matthew.argall@unh.edu}

\begin{document}
\onecolumn
\firstpage{1}

\title[MMS SITL Ground Loop]{MMS SITL Ground Loop: Automating the burst data selection process} 

\author[\firstAuthorLast ]{\Authors} 
\address{} 
\correspondance{} 

\extraAuth{}

\maketitle

\begin{abstract}

\section{}
Global-scale energy flow throughout Earth’s magnetosphere is catalyzed by processes that occur at Earth’s magnetopause (MP). Magnetic reconnection is one process responsible for solar wind entry into and global convection within the magnetosphere, and the MP location, orientation, and motion have an impact on the dynamics. Statistical studies that focus on these and other MP phenomena and characteristics inherently require MP identification in their event search criteria, a task that can be automated using machine learning so that more man hours can be spent on research and analysis. We introduce a Long-Short Term Memory (LSTM) Recurrent Neural Network model to detect MP crossings and assist studies of energy transfer into the magnetosphere. As its first application, the LSTM has been implemented into the operational data stream of the Magnetospheric Multiscale (MMS) mission. MMS focuses on the electron diffusion region of reconnection, where electron dynamics break magnetic field lines and plasma is energized. MMS employs automated burst triggers onboard the spacecraft and a Scientist-in-the-Loop (SITL) on the ground to select intervals likely to contain diffusion regions. Only low-resolution survey data is available to the SITL, which is insufficient to resolve electron dynamics. A strategy for the SITL, then, is to select all MP crossings. Of all 219 SITL selections classified as MP crossings during the first five months of model operations, the model predicted 166 (76\%) of them, and of all 360 model predictions, 257 (71\%) were selected by the SITL. Most predictions that were not classified as MP crossings by the SITL were still MP-like, in that the intervals contained mixed magnetosheath and magnetospheric plasmas. The LSTM model and its predictions are public to ease the burden of arduous event searches involving the MP, including those for EDRs. For MMS, this helps free up mission operation costs by consolidating manual classification processes into automated routines.

\tiny
 \keyFont{ \section{Keywords:} Magnetospheric Multiscale (MMS), Scientist in the Loop (SITL), Burst Data Management, Magnetopause, Long-Short Term Memory (LSTM), Ground Loop} 
\end{abstract}


\section{Introduction}
\label{sec:intro}
Earth is a strongly magnetized planet whose internal dynamics are largely influenced by its interaction with the solar wind and the resulting cycle of magnetic reconnection \citep{Dungey:1961}. Reconnection occurs initially at the magnetopause (MP), at the interface between the shocked solar wind and Earth's magnetosphere (MSP), in what is known as the electron diffusion region (EDR). The EDR had been enigmatic, with few direct observations \citep{Nagai:2011, Nagai:2013, Oka:2016, Scudder:2012, Tang:2013} because spacecraft lacked the spacial and temporal resolution to resolve electron-scale dynamics. These limitations were overcome by the Magnetospheric Multiscale (MMS) mission. Leading up to its launch, so little was known about the EDR that it was unclear how or if EDRs could be identified in the data, which led to speculation into the best EDR indicator \citep{Mozer:2005a, Scudder:2008a, Scudder:2008b, Zenitani:2011, Swisdak:2016, Aunai:2013b, Hesse:2014b}. Since launch, however, MMS has identified more than 50 EDRs (see \citet{Webster:2018} for a partial list) and greatly expanded our knowledge of what catalyzes the global reconnection cycle.


To do this, MMS made significant efforts to capture enough of the right data to achieve its mission goals. The amount of high time resolution burst data recorded onboard is such that only about 4\% can be downlinked. We present the first machine learning (ML) model that has been fully implemented into the mission operation's data flow in order to ensure that the selected 4\% of burst data is able to address MMS science objectives \citep{Burch:2016a, Torbert:2016a}. Our efforts help to transfer mission operations resources to science and to work around external constraints to advance our understanding of energy flow within the MSP.

MMS is not the only mission to face data limitations. Missions such as WIND, THEMIS, Cluster, STEREO, and others have burst mode schemes that operate only when triggered. Some burst modes are triggered on a pre-determined duty cycle, while others are triggered on an instrument-by-instrument basis when on board measurements meet certain criteria. Still others are coordinated among multiple or all instruments. While burst modes and their triggers are mentioned in the instrumentation literature, details about the algorithms and the criteria behind them are mostly omitted, and their efficacy is largely unknown. Some WIND and THEMIS triggers used to detect plasma boundaries are described by \citet{Phan:2015}. Triggers used on STEREO for shock detection, their evolution, and their efficacy are documented in \citep{Jian:2013}. The MMS mission, driven by the large data volumes required to study electron dynamics at Earth’s MP, is the first to fully document its burst system from the beginning of the mission. In this paper, we describe our efforts to build upon the early mission design work in order to automate the burst data selection process.

The MMS burst management system consists of the automated burst system (ABS) that selects burst intervals by passing 10\,s averaged trigger data numbers (TDNs) to on-board tables that set the burst trigger criteria \citep{Fuselier:2016,Baker:2016}, and a human Scientist-in-the-Loop (SITL) who examines all of the low-resolution survey data, and who manually selects and classifies burst intervals. Survey data, however, is insufficient to resolve electron dynamics. A strategy for the SITL, then, is to select all MP crossings. This has resulted numerous EDR encounters but is labor- and resource-intensive; after manual reclassification, just $\sim$0.7\% of MP crossings, or $\sim$0.0001\% of the mission lifetime during MMS’s first two years contained an EDR. Such challenges were foreseen when designing the ABS and SITL selection processes and it was envisioned that automated algorithms would supplement or replace them. Algorithms that use the survey data available to the SITL on the ground fit into the ``ground loop''. Presented below are the design, implementation, and results of the first ground loop.


Most applications of ML to magnetospheric physics problems to-date have been geared toward the prediction of catastrophic events, including geomagnetically induced currents \citep{Wintoft:2015} that threaten power grids, solar energetic particles that threaten space assets \citep{Boubrahimi:2017}, and geomagnetic indices \citep{Lundstedt:1997,Bhaskar:2019,Borovsky:2014} that indicate when global geomagnetic activity could lead to such events. Most methods associate upstream conditions at L1 to those at geosynchronous orbit or on the ground because of the continuous data coverage linking upstream and downstream conditions. Unfortunately, because such models lack knowledge of processes internal to the MSP (e.g. \citet{Bhaskar:2019}), they tend to suffer precisely during the extreme events they are trying to predict. Space weather prediction can be improved by creating several ML models with knowledge of specific aspects of magnetospheric dynamics. We later describe how several MMS ground loops could be combined to identify complex geomagnetic processes and support special science campaigns.

The primary science goal of MMS is to study electron dynamics associated with magnetic reconnection. However, because electron dynamics are not resolved in the survey data, the strategy employed by the SITL during the dayside phase is to select all MP crossings, a prominent location for magnetic reconnection \citep{Phan:2015}. Past attempts to identify the MP in an automated fashion used gradients in plasma parameters such as density or ion flux \citep{Boardson:2000, Phan:2015}. Other methods introduce machine learning to identify the MP indirectly by classifying topologically distinct regions then locating the transition between them. The solar wind, magnetosheath, and magnetosphere were identified by applying probability functions \citep{Jelinek:2012} and support vector machines \citep{daSilva:2020} to ion density and temperature data, 3D convolutional neural network to 3D ion distribution functions \citep{Olshevsky:2019}, and random forests to magnetic field and plasma data \citep{Nguyen:2019b}. The MP is then inferred as the boundary between the MSH and MSP. We present the first model specifically trained to identify MP crossings, thereby automating the primary SITL task.

It is clear from the number of MP classifiers above that identifying the MP is important not just for the SITL. The MP is the primary location of mass, momentum, and energy transfer into Earth's magnetosphere. Because of this, many statistical studies have focused on MP properties \cite{Paschmann:1993, Phan:1996}; MP processes like flux transfer events \citep{Fear:2012}, Kelvin-Helmholtz instabilities \citep{Kavosi:2015}, velocity rotation events \citep{Matsui:2019}, impulse events, and kinetic Alfv\`en waves \citep{Wing:2014}; and creating MP models \citep{Jelinek:2012, Boardson:2000}. Statistical studies such as these traditionally require arduous event searches. Automated algorithms and event lists can be used as a first-step data filter to make searches less burdensome. For this reason, our model and its predictions are publicly available for use in future studies \citep{Argall:2020c, Small:2020}.

This paper serves two purposes: 1) to document the burst management system and infrastructure and 2) demonstrate the performance of the first ground loop ML model. It is outlined as follows. First, the systems for making burst selections, including the SITL, ABS, and GLS, are described in \S\ref{sec:burst-selection}. Next, an overview of the tools and processes developed to support the GLS infrastructure is provided in \S\ref{sec:infrastructure-tools}. Then, a description of the data is given in \S\ref{sec:data} and the model in \S\ref{sec:mp-model}. In \S\ref{sec:performance}, we present the model results and performance. Section~\ref{sec:discussion} is the Discussion, and \S\ref{sec:gls-hierarchy} outlines the GLS Hierarchy, a framework needed to fully automate the SITL selection process. Finally, a summary is given in \S\ref{sec:summary}. Those interested in only the model and its results are referred to \S\ref{sec:mp-model} and \S\ref{sec:performance}.

\section{Burst Management Systems}
\label{sec:burst-selection}
MMS burst memory management consists of three systems for selecting intervals of burst data for downlink: the Scientist-in-the-Loop (SITL), the Automated Burst System (ABS), and the Ground Loop System (GLS).

\subsection{Scientist-in-the-Loop}
\label{sec:sitl}
The SITL is a role that rotates among MMS team members. Currently, there are 73 participating SITL scientists that have made selections on over 1090 orbits of data. Each orbit contains Sub-Regions of Interest (SROIs) that encompass the most probable MP location, the bow shock, and other regions of scientific interest. SITLs make selections from the SROIs within a SITL window, a timeframe in which the MMS satellites make contact with ground-based radio communication network and incrementally downlink data. Data is passed through to the Science Data Center (SDC) (\S\ref{sec:sdc}) where preliminary calibrations are applied and the data is made available to the SITL. The SITL then uses the EVA tool (\S\ref{sec:eva}) to interactively select data intervals for downlink.

\begin{table}
    \centering
    \begin{tabular} { p{1.5in} | p{1in} | p{3in} }
    Event Type & FOM Category & Data Signatures / Notes \\
    \hline\hline
    Complete High Magnetic Shear Magnetopause Crossing & 1 (1- if very long crossing or low-shear) & full density gradients and full magnetic field rotations, includes separatrix and exhaust boundary \\
    \hline
    Magnetopause diffusion region candidate & 1+ & Reversals of high speed jet and B normal during magnetopause crossing \\
    \hline
    Magnetopause diffusion region candidate & 1+ & Magnetopause without boundary layer.  At current sheet center: positive sunward pointing normal electric field. (Such events can be difficult to identify in SITL data) \\
    \hline
    Magnetopause: Kelvin-Helmholtz induced current sheet & 2 & Quasi-periodic magnetic field and density oscillations, field direction changes. Can select a long interval (tens of minutes) if spectacular. \\
    \hline
    Magnetopause: FTE & 2 & Bipolar B normal and strong enhancement of $|B|$ \\
    \hline
    Magnetopause: partial crossings & 2 & Incomplete B rotation and density transition (i.e., not reaching magnetospheric levels) \\
    \hline
    Boundary layer traversals & 3- & Excursion into the boundary layer, characterized by magnetosphere B and slight increase in density and appearance of magnetosheath ion population.
    \end{tabular}
    \caption{Guidelines used during Phase 5A (30 Sept. 2019 through 24 Nov. 2019) specifying how the SITL should classify magnetopause crossings. Events in categories 1-4 are given FOMs of 150-199, 100-149, 50-99, and 0-49, respectively. A "+" or "-" after a given category indicate a selection at the upper- or lower-range of the category. FOMs $\ge 200$ are reserved for special events, such as calibration intervals or a definitive EDR encounter.}
    \label{tab:sitl-guidelines}
\end{table}

SITLs follow guidelines set by mission PIs and Super SITLs (SITLs with super-user privileges) to help standardize the selection process. Those related to the MP are provided in Table~\ref{tab:sitl-guidelines}. Each selection is given a Figure of Merit (FOM), a ranking between 0 and 255 split into five categories, to prioritize which selections are downlinked first. Priorities change based on the type of MP crossing. For example, complete, high-shear MP crossings receive a category 1 ranking (FOM 150-199), indirect reconnection signatures such as FTEs receive a category 2 ranking (FOM 100-149), and boundary layer encounters receive a category 3 ranking (FOM 50-99). Using these guidelines, the SITL makes a median of 30 selections per orbit, with a maximum to-date of 200 selections in a single orbit. The time spent by the SITL in making such selections can be saved by automating the process.

\subsection{Automated Burst System}
\label{sec:abs}
The ABS applies configurable tables of weights and offsets to 10\,s averaged burst quantities from each instrument, named Trigger Data Numbers (TDNs), to assign a Cycle Data Quality (CDQ) index to each 10\,s buffer of burst data. The four CDQ values provided by the four spacecraft are downlinked, multiplied by another weighting factor, then summed to provide an overall Mission Data Quality (MDQ) index. The MDQ index is used to prioritize data for download \citep{Fuselier:2016}.

There are 34 TDN terms available to the ABS. Early in the mission, the system was configured to look only for large changes in the magnetic field $B_{z}$ component. Reconnection events identified by scientists during the first two years of the mission have subsequently been used to determine which of the TDNs efficiently parameterize reconnection and to determine their relative importance.

For the dayside magnetopause, a set of 6 parameters is now employed in the search for reconnection. These involve changes in the magnetic field components, the electric field wave power, the electron pressure, and the ion density. The parameters and their corresponding weights were determined based on their ability to select the 32 intervals that contained potential dayside EDRs identified by \citet{Webster:2018}. The ABS as it is now configured would have selected 31 of the 32 \citet{Webster:2018} events for download with efficiency comparable to that of the SITL.

For the magnetotail, a different set of parameters is used. Six trigger terms that respond strongly to reversals of the magnetic field and bulk velocity were identified in data acquired during the 2017 magnetotail phase of the mission.  Weights and gains were optimized through linear regression. The resulting ABS tables currently in use would have captured two well-substantiated magnetotail EDR encounters \citep{Torbert:2018,Zhou:2019} with total download of burst intervals equivalent to the actual number of SITL selections.

\subsection{Ground Loop System}
\label{sec:gls}
The GLS is designed to be a system of ML or empirical models that automate the event classification process using all of the data available to the SITL. Data available to the SITL is of restricted use because it is lower quality that the Level-2 science-quality data freely available to the public. ML models trained on SITL data, such as the MP model described in this manuscript, may not perform as well when applied to Level-2 data. And, vice versa, a model trained using Level-2 data may under-perform if incorporated into the GLS.

The first GLS model (\S\ref{sec:mp-model}) uses the text description given to each burst selection by the SITL as the ground-truth manual classification for training purposes. To encourage expansion of the GLS, our model development notebooks \citep{Small:2020} can generate additional models simply by changing the text filter (e.g. replace ``Magnetopause'' with ``Dipolarization Front''). SITL classifications significantly reduce the time required to train a supervised learning model, and the variety of selections made can facilitate a hierarchical ground loop infrastructure (\S\ref{sec:gls-hierarchy}), thereby reducing the burden of the SITL and allowing them to spend more time looking for new science.

\section{Infrastructure and Tools}
\label{sec:infrastructure-tools}

\subsection{Science Data Center}
\label{sec:sdc}
The MMS Science Data Center (SDC) is a collection of virtual machines and software applications that collectively support the science data processing and data access requirements for the MMS mission. It has been running since mission launch in March 2015 and currently manages a collection of over 11 million science data files accessible to MMS mission team members and 4 million files available to the public.

One of the key activities for the GLS is the ability to process the data used as input to the ground loop prediction models. This activity starts with a fixed time schedule or an external event set to trigger a science data processing job. The event sets relevant to the GLS are Deep Space Network (DSN) contacts that transfer spacecraft telemetry data to a ground station. The ground station transfers the raw data files to a NASA facility which then uploads them to the LASP Payload Operations Center (POC), where they are ingested to a raw telemetry database. A spacecraft-specific processing task is scheduled at the end of each DSN contact, delayed enough to allow the various data transfer and ingest tasks to complete. Each MMS instrument has a set of associated processing tasks for different data rates (survey vs. burst) and data levels. Processing tasks for the SITL ground loop are associated with survey data and the lowest data levels.

The GLS model-evaluation task is delayed an additional amount to allow completion of the various science data products that it needs to evaluate the model. Each GLS job produces a $csv$ file containing the time range and FOM for each of the automated selections. A dropbox manager transfers the ground-loop selections into main SDC storage and indexes it for web-service access by the remote scientist's EVA tool (\S\ref{sec:eva}). The EVA tool can then plot the ground-loop selections alongside those of the ABS and the science data products, allowing the SITL to make informed selections.

The SITL must make selections within 12 hours of observation time to ensure that spacecraft commands used to ``lock'' the selected memory buffers are received before valuable observations are overwritten by newer ones. Although current MMS orbit periods are about 84 hours, spacecraft memory can hold only about 48 hours of burst data. The spacecraft contacts have variable schedules and cannot be optimized just for MMS. The SDC completes the GLS processing within about two hours of the end of each DSN contact, well within required time limits.

The SDC mails reports to a broad team of experienced MMS SITL scientists, allowing review and comment on the latest selections. If it becomes practical to make fully automated selections based on algorithmic analysis, the SDC could short-circuit the human loop and transfer GLS results directly to the POC without human intervention.

\subsection{EVA}
\label{sec:eva}
EVA is a graphical user interface (GUI) designed specifically for the MMS/SITL activity and provided as a part of the MMS plug-in for the Space Physics Environment Data Analysis Software (SPEDAS) package \citep{Angelopoulos:2019}. SPEDAS is a software package for the IDL language that provides scripts for convenient plotting of spacecraft time series data and particle distributions. The MMS plugin includes the EVA GUI software, as well as software routines to load and plot data from every instrument onboard MMS.  The main functions of EVA are to:

\begin{enumerate}
  \item Load and display reduced-resolution, survey data for the entire duration of Region-of-Interest (ROI)
  \item Help the SITL to identify and prioritize scientifically-valuable time ranges for downlinking the full-resolution burst data, and 
  \item Send the list of selected time ranges and their FOM values back to SDC for commanding.
\end{enumerate}

A key feature of EVA is that it provides some pre-defined parameter sets. Parameter sets consist of data products frequently used by the SITLs (many are listed in Table~\ref{tab:features}) in combinations tailored to individual instruments, specific investigations, or regions of space. For example, there are parameter sets for the magnetopause, magnetotail, bow shock, and solar wind. By selecting a specific parameter set, in addition to the desired spacecraft ID, date, and time period, a SITL scientist can go straight to the task of viewing data needed for the SITL activity. 

Parameter sets are displayed as tiles of time-series data in an interactive window in which the SITL can add, edit, and delete burst selections. In the earlier phases of the mission, EVA would append an ABS selections panel to the bottom of the window and the SITL scientist would manually adjust them while inspecting the data. Today, ground loop selections are also available to better guide the SITL in their selection process. This helps reduce personal bias when selecting similar phenomena, such as MP crossings.

\subsection{PyMMS}
\label{sec:pymms}
PyMMS \citep{Argall:2020a} is a software package written in Python and freely available on GitHub and PyPI that makes full use of the MMS SDC's data API. It is able to download instrument data (including both SITL and L2 quality data), as well as the ABS, GLS, and SITL selections. The SITL provides an ASCII text description of each burst interval that they select, which can be easily downloaded and searched with PyMMS to train supervised learning models. The GLS model described in this paper (\S\ref{sec:mp-model}) was trained in this way.


Note that the SITL, GLS, or ABS selections can be submitted multiple times (\S\ref{sec:sdc}), often with changes, so the available selections files have duplicate and overlapping entries, and may not necessarily be in chronological order. Also, because of downlink and storage limitations, selections of long duration are broken into smaller chunks. PyMMS has tools to deal with these issues.

\section{Data}
\label{sec:data}

\begin{figure}[b]
	\centering
	\includegraphics[width=0.8\linewidth]{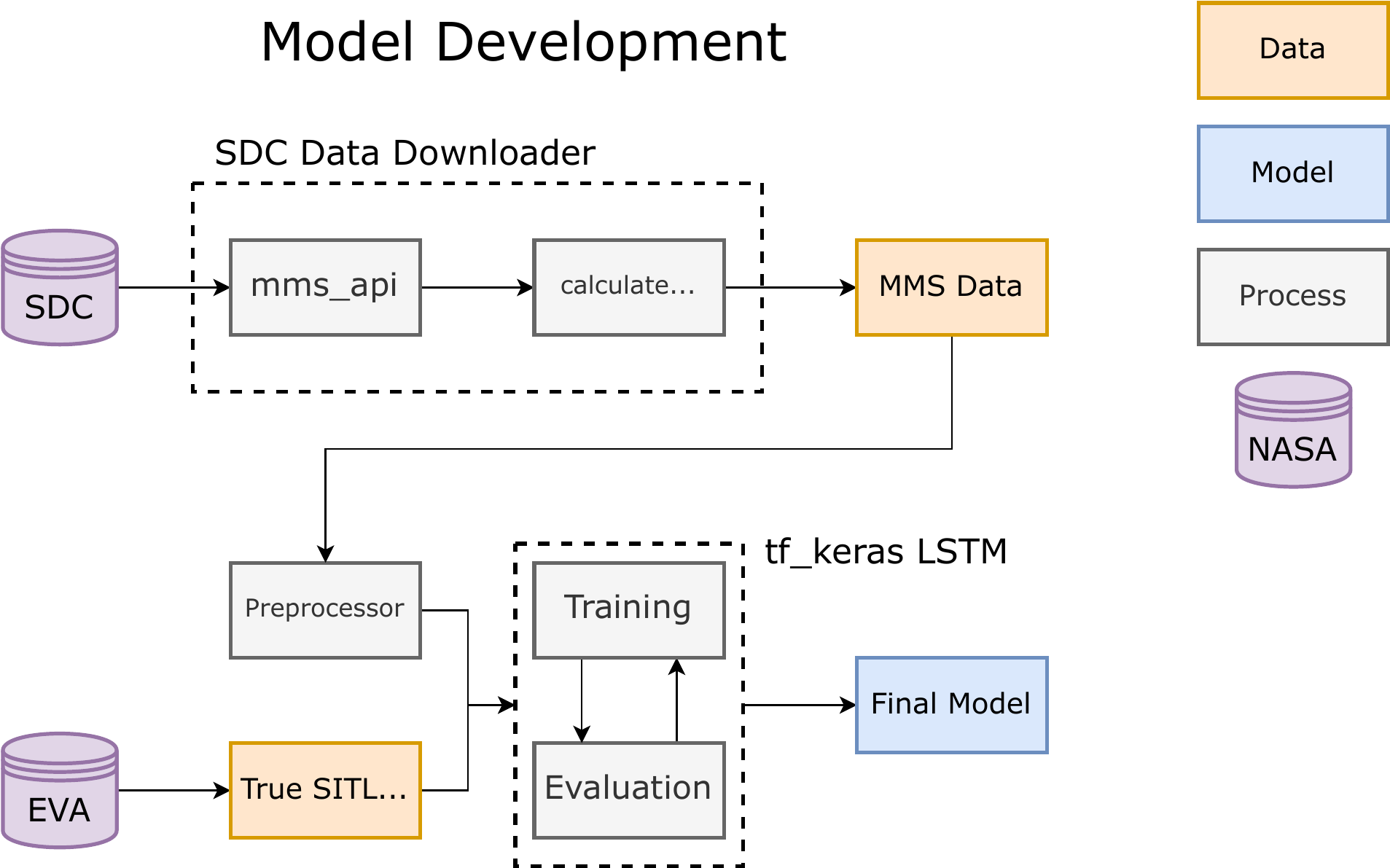}
	\caption{The GLS magnetopause model is an example of a supervised learning model that made use of data labeled previously by the SITL for training and testing. Applying historical SITL labels to preprocessed data significantly reduce the time needed for model development.}
	\label{fig:model_flow}
\end{figure}

The flow of data used to train the MP model is shown in Figure~\ref{fig:model_flow}. The SDC provides data from the Analog Fluxgate (AFG) magnetometer \citep{Russell:2014}, the Electric field Double Probes (EDP) \citep{Ergun:2014,Lindqvist:2014}, and the Fast Plasma Investigation (FPI) Dual Ion Spectrometer (DIS) and Dual Electron Spectrometer (DES) \citep{Pollock:2016}. Fast survey data from each instrument was subjected to a preliminary set of calibrations to produce SITL-quality data, which is suitable for making informed decisions about burst selections but not for deep scientific scrutiny. This is necessary because of the urgency with which selections need to be made.

From these products, the 123 features listed in Table~\ref{tab:features} were chosen to be inputs into the ML model. Most are standard products from the instruments, such as the $\mathbf{B}$ and $\mathbf{E}$ fields and their magnitudes, and the plasma energy spectrograms and moments. Others, like the temperature anisotropy, the custom $\gamma$ values, and the $Q_{\Delta}$ values are metafeatures, features computed from the standard features. The $Q_{\Delta}$ features are gradient-based trigger terms used on Wind and THEMIS, and are calculated as $Q_{\Delta x} = |x - \bar{x}|$, where $\bar{x}_{j+1} = [\bar{x}_{j}(2^{M} - 1) + x_{j}]/2^{M}$, and $M=2$ sets the amount of smoothing \citep{Phan:2016}. This set of features represents a large portion of the data available to the SITL.

It is worth noting that all features listed in Table~\ref{tab:features} are calculated from single spacecraft data; in this case, MMS1. This was done primarily for three reasons: 1) contact times for data downlink are variable so data from multiple spacecraft is not guaranteed, 2) orbit configurations can change; spatial gradients valid in a tetrahedron configuration are not be valid in a strong of pearls, and 3) events may occur at different times for different spacecraft, especially in a string of pearls. If the model does not produce satisfactory results for other spacecraft, or if an instrument on MMS1 experiences a problem that invalidates any of the features in Table~\ref{tab:features}, the model can be retrained.

\begin{table}
    \centering
    \begin{tabular}{ l l l p{3in} }
            & Instr. & Feature & Description \\ \hline
            & AFG & & \\ \cline{2-4}
    1-3     & & ($B_{x}$, $B_{y}$, $B_{z}$)                   & X, Y, and Z-components of the magnetic field in DMPA coordinates \\
    4       & & $|B|=\sqrt{B_{x}^{2} + B_{y}^{2} + B_{z}^{2}}$ & Magnitude of the magnetic field     \\
    5       & & $Q_{\Delta B_{x}}$                             & Quality value for $B_{x}$            \\
    6       & & $P_{B} = |B|^{2}/(\mu_{0})$                    & Magnetic pressure                    \\
    7       & & $\theta_{C} = \arctan{(B_{y}/B_{x})}$         & Clock angle.                         \\
            & EDP & & \\ \cline{2-4}
    8-10    & & ($E_{x}$, $E_{y}$, $E_{z}$)                   & X-, Y-, and Z-component of the DC electric field in DSL coordinates \\
    11      & & $|E| = \sqrt{E_{x}^{2} + E_{y}^{2} + E_{z}^{2}}$  & Magnitude of the electric field      \\
            & DIS & & \\ \cline{2-4}
    12-42   & & $\mathcal{E}_{i}$                             & Omni-directional energy spectrogram  \\
    43      & & $N_{i}$                                       & Number density                       \\
    44-45   & & ($V_{i,x}$, $V_{i,y}$)                        & Bulk velocity in DBCS coordinates    \\
    46-47   & & ($Q_{i,xx}$, $Q_{i,yy}$)                      & Heat-flux vector in DBCS coordinates \\
    48-49   & & $T_{i,\parallel}$, $T_{i,\perp}$              & Parallel and perpendicular temperatures \\
    50-55   & & $\underbar{P}_{i}$                            & Upper diagonal elements of pressure tensor in DBCS coordinates \\
    55-60   & & $\underbar{P}_{i}$                            & Upper diagonal elements of pressure tensor in DBCS coordinates \\
    61      & & $A_{i} = T_{i,\parallel}/T_{i,\perp} - 1$     & Temperature anisotropy               \\
    62      & & $T_{i} = (T_{i,\parallel} + 2 T_{i,\perp})/3$ & Scalar temperature                   \\
    63      & & $Q_{\Delta N_{i}}$                            & Quality value for $N_{i}$            \\
    64      & & $Q_{\Delta V_{i,z}}$.                         & Quality value for the $V_{i,z}$      \\
    65      & & $Q_{\Delta N_{i} |V_{i}|}$.                    & Quality value for ion ram pressure   \\
            & DES & & \\ \cline{2-4}
    66-97   & & $\mathcal{E}_{e}$                             & Omni-directional energy spectrogram  \\
    98      & & $N_{e}$                                       & Number density                       \\
    99-100  & & ($V_{e,x}$, $V_{e,y}$)                        & Bulk velocity in DBCS coordinates    \\
    101-102 & & ($Q_{e,xx}$, $Q_{e,yy}$)                      & Heat-flux vector in DBCS coordinates \\
    103-104 & & $T_{e,\parallel}$, $T_{e,\perp}$              & Parallel and perpendicular temperatures \\
    105-110 & & $\underbar{P}_{e}$                            & Upper diagonal components of pressure tensor in DBCS coordinates \\
    111-116 & & $\underbar{P}_{e}$                            & Upper diagonal components of pressure tensor in DBCS coordinates \\
    117     & & $A_{e} = T_{e,\parallel}/T_{e,\perp} - 1$     & Temperature anisotropy               \\
    118     & & $T_{e} = (T_{e,\parallel} + 2 T_{e,\perp})/3$ & Scalar temperature                   \\
    119     & & $p_{e} = (P_{e,xx} + P_{e,yy} + P_{e,zz})/3$  & Scalar pressure                      \\
    120     & & $Q_{\Delta N_{e}}$                            & Quality value for $N_{e}$            \\
    121     & & $Q_{\Delta V_{e,z}}$                          & Quality value for $V_{e,z}$ velocity \\
            & Multiple & & \\ \cline{2-4}
    122     & & $\gamma_{1} = T_{i} / T_{i}$                  & Custom feature                       \\
    123     & & $\gamma_{2} = 2 T_{i} / |E|$                  & Custom feature
    \end{tabular}
    \caption{Features used for the development and application of the GLS magnetopause model. Post-processing on the ground allows the raw data to be partially calibrated and expanded into a richer dataset than is available on the spacecraft. The model makes use of most data available to the SITL plus metafeatures that were used as burst triggers for previous missions.}
    \label{tab:features}
\end{table}

Data from 1 Jan. 2017 to 30 Jan. 2017 were used to train the model. During this time period, MMS had a single SROI keyed on apogee ($8 < X_{GSE} < 12\,R_{E}$); apogee was at $12\,R_{E}$ geocentric distance and was located near the subsolar point ($-6 < Y_{GSE} < 0\,R_{E}$). The amount of training data was limited by resources on the platform on which model training was performed. Once downloaded, all data outside of the intervals selected and labeled by the SITL were discarded to ensure the accuracy of the ground truth. The data was then interpolated onto the 4.5\,s cadence of the FPI fast survey data products, then scaled and regularized. Because of an imbalance between the number of measurements that were selected as MP crossings compared to those that were not, class weights were applied to the ``MP'' and ``not MP'' classified data. We normalized all features with standardization, calculated as $x' = \frac{x - \bar{x}}{\sigma_{x}}$, where $x$ is the original vector for a given feature, $\bar{x}$ is the average of that vector, and $\sigma_{x}$ is its standard deviation. This method of normalization is widely used in machine learning applications to boost the performance of the model's gradient descent while learning. Finally, each orbit was broken down into consecutive sequences of 250 measurements to reduce computational complexity. Of all such sequences, 80\% were used for training and 20\% were used for testing.

\section{GLS Magnetopause Model}
\label{sec:mp-model}



We develop an ML model that aids and automates a key task performed by the SITL. Using the same low-resolution data as the SITLs, we identify time intervals that are likely to contain MP crossings. The ML models are trained using historical data annotated by SITL selections. The input is a low-resolution (4.5\,s) time sequence of the data quantities outlined in Table~\ref{tab:features}. The goal is to predict, for each data point at time $t$, whether the particular 4.5\,s interval would be selected by the SITL as an MP event. 

\begin{figure}
    \centering
    \includegraphics[width=\linewidth]{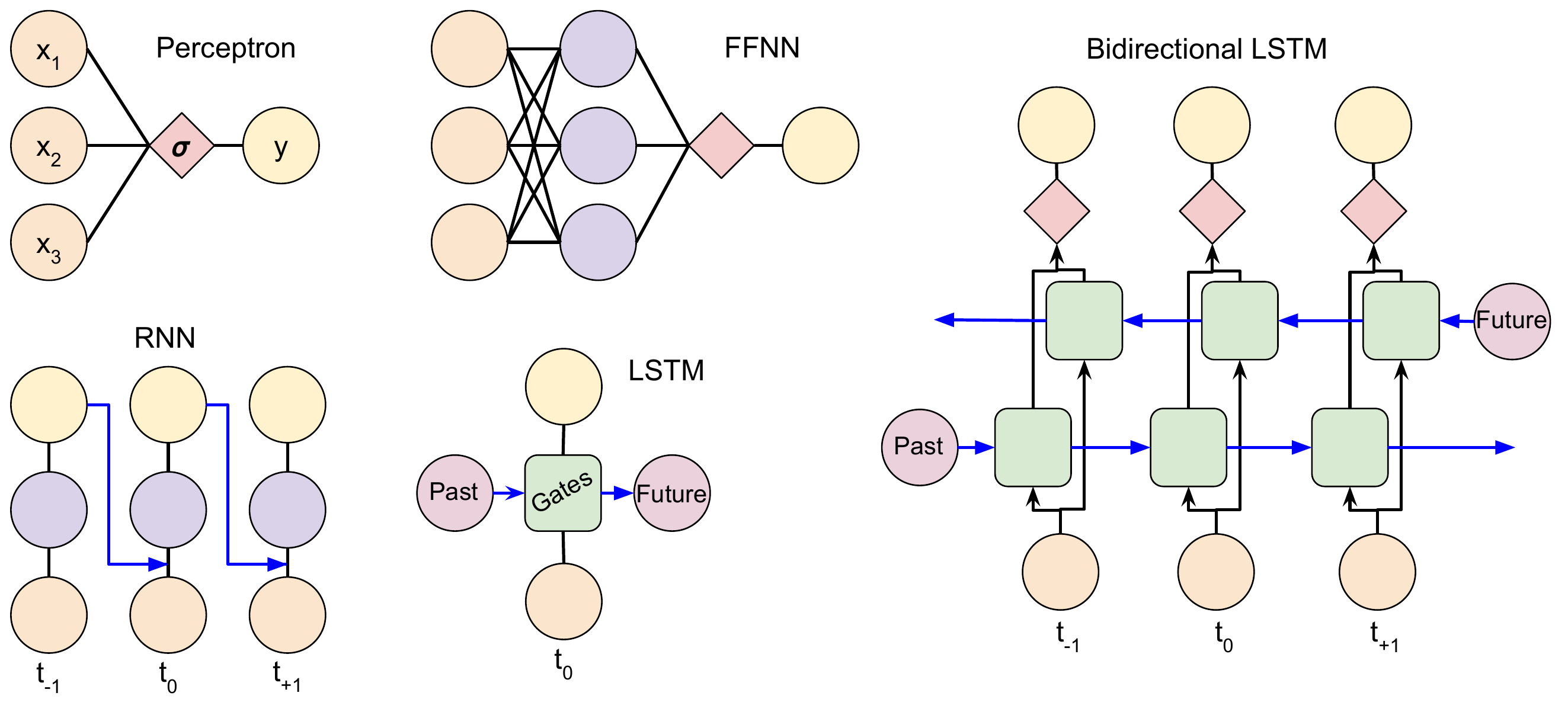}
    \caption{Evolution of the non-linearity and time-dependence of neural network models that make up the bidirectional Long Short Term Memory GLS magnetopause model.}
    \label{fig:neural-nets}
\end{figure}

Our machine learning model is based on neural networks (NNs) \cite{Goodfellow:2016}. A conceptual understanding of the MP model is built up from the NNs shown in Figure~\ref{fig:neural-nets}. The most simple of NNs is a perceptron, which takes a linear combination of the input features, $x_{i}$, for a given time sample and passes the results through a sigmoid (``S''-shaped) activation function that maps the result to the interval $[0,1]$ to predict the output. Training occurs via backpropagation, a process by which the error in the prediction is used to adjust the weights applied to the inputs, usually via some gradient of the sigmoid function. After iterating, the NN output, $y$, converges to a ``yes'' or ``no'' prediction. Perceptrons identify linear relationships between inputs and outputs.

Feed Forward NNs (FFNNs), Recurrent Neural Networks (RNNs), and Long-Short Term Memory (LSTM) Neural Networks evolve the perceptron to learn more complex, non-linear concepts. FFNNs do so by adding hidden layers whose weights determine the relationship between the input features themselves. They are trained via backpropagation in the same way as perceptrons. Like perceptrons, FFNNs make predictions using data from a single time sample. RNNs take the output of a FFNN at time $t_{-1}$ and combine it with the input of the FFNN at time $t_{0}$, thereby incorporating the context inherent to time series data. Training backpropagates errors not only through the hidden layers, but also through time. Vanishing gradients in long prediction chains cause RNNs to have short-term memory. LSTMs create long-term memory by applying gates to information carried forward from past predictions. Our model uses an adaptation of the LSTM to identify MP crossings.

\begin{figure}
    \centering
    \includegraphics[width=0.5\linewidth]{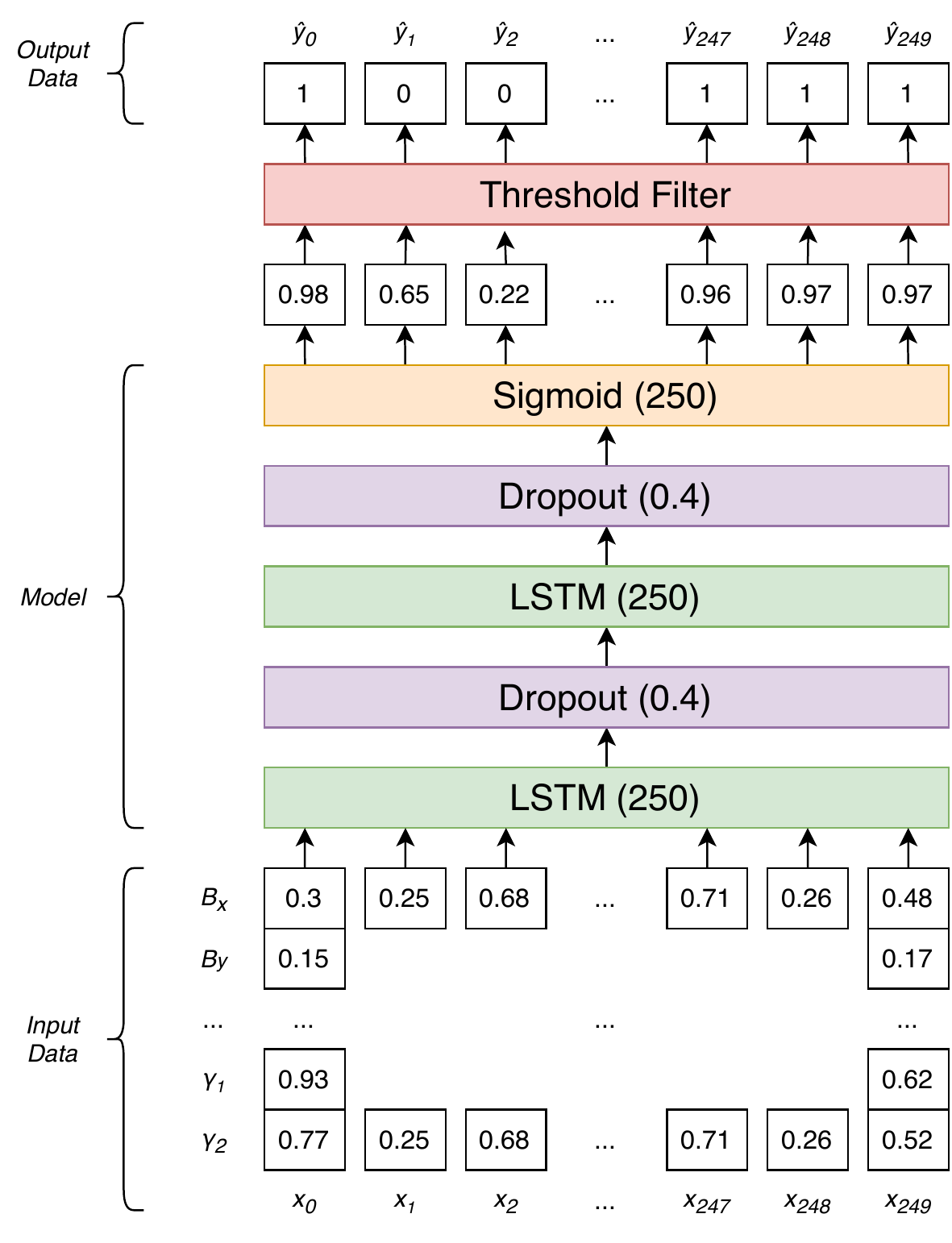}
    \caption{The Architecture of the GLS magnetopause model mimicks how scientists interpret data. It makes decisions by placing data in the context of past and future observations.}
    \label{fig:architecture}
\end{figure}

The GLS MP model is composed of two bidirectional LSTM layers~\citep{Goodfellow:2016}, as depicted in Figure~\ref{fig:architecture}. The output activation functions are hyperbolic tangents and the recurrent activation functions passed to units in $t_{+1}$ and $t_{-1}$ time steps are logistic. Each LSTM layer is followed by a drop-out layer with a drop probability of $0.4$ as a means of forgetting information. Dropout layers help to reduce over-fitting when training the network~\cite{Srivastava:2014}. The output layer is a single unit with a logistic (sigmoid) activation function. The LSTM's output is passed through a threshold filter before contiguous segments of selected points are grouped to form selections.

Contiguous data points with positive predictions are combined to determine the time interval and duration of a suspected MP crossing. These suspected MP crossings are then presented to SITLs during their selection process to quicken, improve, and ultimately replace the manual selection process. Figure~\ref{fig:model_flow} shows a graphical representation of the data flow in our proposed automated SITL model. Data is downloaded from the SDC using PyMMS (\S\ref{sec:pymms}) and is pre-processed (\S\ref{sec:data}) before being fed to our model to identify predicted MP crossings. These predictions are saved to csv files and stored on the SDC's servers until finally transferred to the EVA team for a SITL to view when making selections.

The training and validation data, notebook used to create the model, as well as the weights, scaling parameters, and notebook to run the model are publicly available \citep{Argall:2020c, Small:2020}. Both notebooks have a flag to switch between SITL-quality and science-quality data, and the model creation notebook can easily be modified to generate a new model for any SITL-classified event type. Additional details about the model, threshold filter, and the hardware the model was trained on can be found in the supplemental information.



\section{Results}
\label{sec:performance}
\subsection{Case Studies}

\begin{figure}[t]
    \centering
    \includegraphics[width=\textwidth]{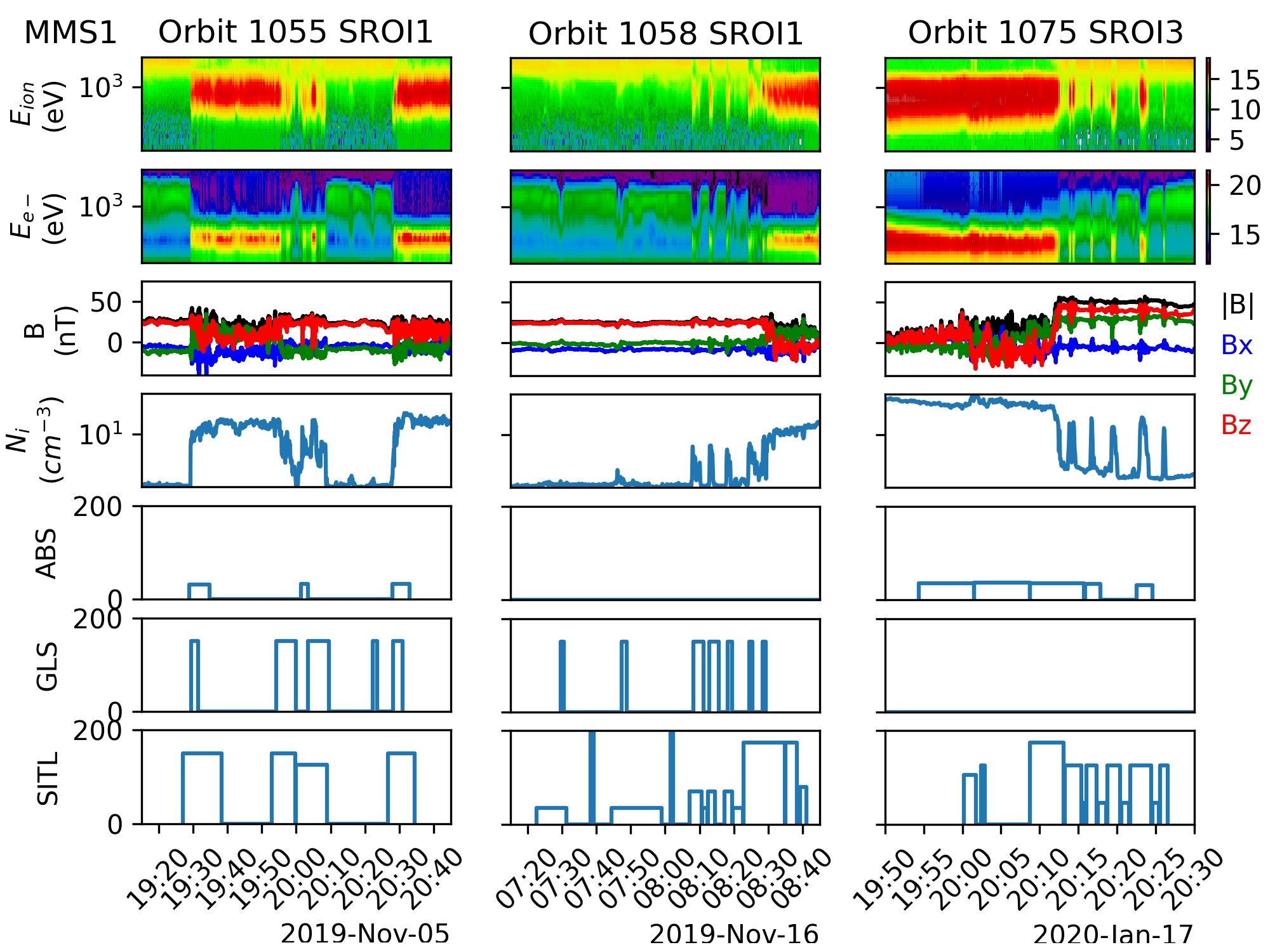}
    \caption{Burst selections made by three different SITL scientists surrounding the magnetopause crossings of three separate orbits. Just as the SITLs have different opinions on how and what to select, and with what FOM value, so too do the GLS and ABS.}
    \label{fig:orbit-performance}
\end{figure}

On 19 Oct. 2019, the model was installed and executed at the SDC for the first time and has been providing guidance to the SITL ever since. By this time, MMS apogee was near $25\,R_{E}$ geocentric distance and had transitioned from one SROI to three SROIs to capture the inbound and outbound magnetopause crossings (SROI1 and SROI3), and a segment of the solar wind (SROI2). Figure~\ref{fig:orbit-performance} shows MP crossings from SROI1 on orbits 1055 (left) and 1058 (center), and SROI3 on orbit 1075 (right), along with the selections made by the ABS, GLS, and SITL. The MP is identified in the ion and electron energy spectrograms (panels 1 and 2) as the transition between the hot, tenuous plasma of the magnetosphere and the colder, denser plasma of the magnetosheath. MMS transitions outward from the MSP to the MSH in SROI1 and inward from the MSH to the MSP in SROI3. During the transition, the MSH and MSP plasmas are observed simultaneously, $B_{z}$ (panel 3) often changes sign, and the density (panel 4) transitions from ${\sim}1\,\mathrm{cm^{-3}}$ in the MSP to $>10\,\mathrm{cm^{-3}}$ in the MSH. SITL selections for these intervals are displayed in the bottom panel while the SITL-provided description of each selection is given in Table~\ref{tab:sitl-selections}. Panels 5 and 6 show similarities and differences between the time intervals and FOM values of selections made by the SITL and those made by the ABS and GLS.

\begin{table}
    \centering
    \begin{tabular} { c c c c c p{3in} }
    \hline
    Orbit & Date & Start & End & FOM & Description \\
    \hline\hline
    1055 & 2019-11-05 & 19:27:03 & 19:38:23 & 150 & Lower shear full magnetopause crossing with Vz flow reversal \\
    1055 & 2019-11-05 & 19:52:53 & 19:59:43 & 150 & Full lower shear magnetopause crossing \\
    1055 & 2019-11-05 & 19:59:53 & 20:09:03 & 125 & Partial magnetopause crossings with deep B-minima \\
    1055 & 2019-11-05 & 20:26:43 & 20:34:33 & 150 & Full low-shear \\
    \hline
    1058 & 2019-11-16 & 07:22:33 & 07:31:23 &  35 & Cold ions \\
    1058 & 2019-11-16 & 07:38:23 & 07:39:13 & 200 & FPI Burst Cal - Segment 2 (1058) - H2 - MSP \\
    1058 & 2019-11-16 & 07:44:23 & 07:59:03 &  35 & Boundary layer \\
    1058 & 2019-11-16 & 08:01:33 & 08:02:23 & 200 & FPI Burst Cal - Segment 3 (1058) - H2 - MSP \\
    1058 & 2019-11-16 & 08:07:13 & 08:10:43 &  70 & BL Traversal \\
    1058 & 2019-11-16 & 08:10:53 & 08:12:23 &  35 & Additional context between BL traversals \\
    1058 & 2019-11-16 & 08:12:33 & 08:14:33 &  70 & BL Traversal \\
    1058 & 2019-11-16 & 08:17:13 & 08:19:33 &  70 & BL Traversal \\
    1058 & 2019-11-16 & 08:19:43 & 08:22:43 &  35 & Additional context between boundary layer traversal and MP Crossing \\
    1058 & 2019-11-16 & 08:22:53 & 08:34:53 & 175 & Full MP potentially with a jet. \\
    1058 & 2019-11-16 & 08:35:03 & 08:38:23 & 175 & Continuation of FULL MP \\
    1058 & 2019-11-16 & 08:38:33 & 08:39:03 &  35 & Additional context between MP and FTE \\
    1058 & 2019-11-16 & 08:39:13 & 08:41:03 &  80 & FTE \\
    \hline
    1075 & 2020-01-17 & 20:00:13 & 20:01:43 & 105 & Magnetosheath IMF rotation with bifurcated signature - unresolved exhaust \\
    1075 & 2020-01-17 & 20:02:23 & 20:02:53 & 125 & Potential magnetosheath flux rope \\
    1075 & 2020-01-17 & 20:08:43 & 20:13:03 & 175 & High-shear complete MP \\
    1075 & 2020-01-17 & 20:13:13 & 20:15:23 & 125 & Partial MPs with $Vz<0$ jetting \\
    1075 & 2020-01-17 & 20:15:33 & 20:15:53 &  45 & Fill \\
    1075 & 2020-01-17 & 20:16:03 & 20:17:23 & 125 & Partial MPs \\
    1075 & 2020-01-17 & 20:17:33 & 20:18:33 &  45 & Fill \\
    1075 & 2020-01-17 & 20:18:43 & 20:20:23 & 125 & Partial MPs \\
    1075 & 2020-01-17 & 20:20:33 & 20:21:33 &  45 & Fill \\
    1075 & 2020-01-17 & 20:21:43 & 20:24:23 & 125 & Partial MPs \\
    1075 & 2020-01-17 & 20:24:33 & 20:25:23 &  45 & Fill \\
    1075 & 2020-01-17 & 20:25:33 & 20:26:33 & 125 & Partial MPs \\
    \hline
    \end{tabular}
    \caption{Selections made by the SITL during the intervals shown in Figure~\ref{fig:orbit-performance}. ``Partial'', ``low-shear'', ``high-shear'', ``full'', and ``complete'' refer to classes of MP crossings that receive different FOM values. By selecting a ``fill'' interval at low-FOM, the SITL provides context to adjacent events that is saved on board and can be increased to higher FOM later by a Super-SITL. ``FPI Burst Cal'' = calibration, ``FTE'' = flux transfer event.}
    \label{tab:sitl-selections}
\end{table}

Orbits 1055 and 1058 SROI1, and orbit 1075 SROI3 were chosen because of the presence or absence of GLS and ABS selections. Orbit 1055 SROI1 consisted of three full and one partial MP crossings as the MP moved back and forth over the spacecraft during the 1.5 hour interval shown. Both the ABS and GLS made selections similar to the SITL, but with notable differences. First, whereas the SITL selected a large portion of the MSP and MSH on either side of the MP to provide relevant contextual information, the ABS and GLS selections were more focused on the MP transition. In this case, they are under-selecting when compared to the SITL, but are correctly identifying the MP (i.e. they are not classifying the surrounding MSP and MSH as the MP like the SITL did). Second, the GLS makes a selection at ${\sim}$2022 that the SITL does not. This selection represents a typical false-positive for the GLS despite the fact that the interval is MP-like in that the MSP and MSH plasmas are observed simultaneously and the plasma density is higher than the MSP proper. So, while differences exist between the SITL, ABS, and GLS, those differences are often subtle, and similar differences also exist between selections made by different SITL scientists. We consider the SITL to be the ground truth for the sake of model validation; however, operationally the GLS acts as a co-SITL with its own suggestions for what should be selected.

During orbit 1058 SROI1, the SITL selected intervals containing cold ions (${\sim}$0722-0722), the boundary layer (${\sim}$0744-0820), an MP crossing (${\sim}$0823-0845), and a flux transfer event (${\sim}$0839-0841; see Tab.~\ref{tab:sitl-selections}). None of these intervals were selected by the ABS but the GLS captured most of them. In particular, the intervals marked ``cold ions'' and ``boundary layer''/``BL'' were selected by the GLS because they show signs of mixing of MSP and MSH plasmas, as occurs at the MP. Such selections are similar in nature to the false positive shown in orbit 1055 SROI1.

Orbit 1075 SROI3 is one in which the SITL and ABS select the MP but the GLS does not. The SITL selects a complete, high-shear MP crossing at ~2010 UT followed by several partial crossings (Table~\ref{tab:sitl-selections}), as well as reconnection-like signatures in the MSH near 2000 UT. The ABS also selects the high-shear crossing but only some of the partial crossings. It also captures the reconnection signatures in the MSH. For this time interval, the ABS over-selects in the MSH and under-selects at the MP. As for the GLS, further testing on this interval reveals that the GLS does select the MP if the LSTM model is run on a limited interval surrounding the MP, as opposed to the entire SROI. More generally, the GLS selects the majority of MP crossings during all SROI1 intervals but very rarely selects MP crossings during SROI3. These two facts could indicate that the training and validation sets need to be expanded to include data from a time period when MMS had three SROIs.


\subsection{Statistical Study}
\begin{figure}[t]
    \centering
    \includegraphics[width=0.7\textwidth]{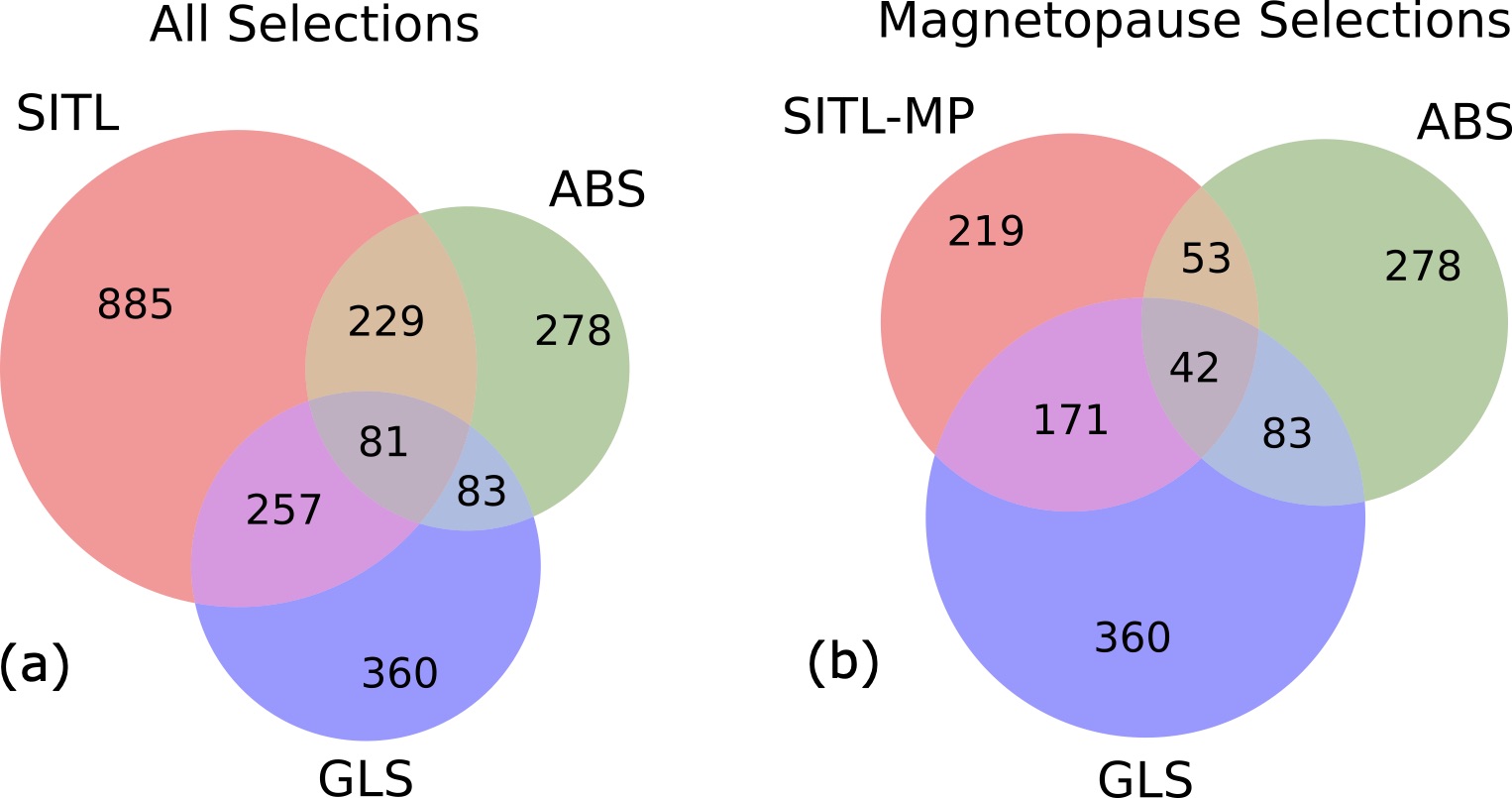}
    \caption{A comparison of SITL, GLS, and ABS segments from SROI1 that takes into account the range-based nature of event selection intrinsic to time series data. }
    \label{fig:venn-diagram}
\end{figure}

To more broadly assess model performance, we make comparisons between the GLS, SITL, and ABS for all selections made in SROI1 between 19 Oct. 2019 and 25 March 2020. Figure~\ref{fig:venn-diagram} is a Venn diagram depicting (a) the number of GLS and ABS selections that have at least partial overlap with all SITL selections and (b) only those SITL selections that were identified as MP crossings. More detailed histograms showing the two-way overlap between between SITL and GLS, SITL and ABS, and GLS and ABS are included in the supplementary material as Figures~S1-3 for SROI1, SROI3, and SROIs1 and 3, respectively. Such comparisons take into account partial- and multiple-overlaps between SITL and GLS selections, something not possible with the more traditional precision, recall, and F1-score metrics \citep{Tatbul:2018} presented in \S\ref{sec:model-scores}

\subsubsection{SITL-GLS Comparison}
Most selections made by the GLS are of interest to the SITL, but the GLS is selecting more than just MP crossings. Of the 360 GLS selections, 257 (71\%) were also selected by the SITL and 176 (49\%) were classified as MP crossings by the SITL. At first glance, the latter might seem low for a model that was trained to select MP crossings. This can be explained partly because the SITL is aware of external control factors such as telemetry restrictions, and partly because some GLS segments are MP-like but are not classified as MP crossings by the SITL, as in Figures~\ref{fig:orbit-performance}a,b. Such selections include intervals that exhibit plasma mixing between the MSP and MSH. They also include bow shock crossings, which have field and plasma gradients similar to those present during MP crossings. While not MP crossings, these extra selections made by the GLS are still of interest to the SITL, as indicated by the larger number of overlapping segments when all of the SITL selections are included.

Again, because of the possibility of partial and multiple overlaps in time series selections, the inverse relationship relating the number SITL selection also selected by the GLS can only be qualitatively inferred (for a quantitative comparison, see Figs.S1-3). On one hand, the GLS selections represent only a small percentage (30\%) of all SITL selections. On the other hand, the majority of SITL MP crossings (78\%) are selected by the GLS. Machine learning models, therefore, are an effective means of automating subsets of SITL functions outlined in the SITL guidelines. Later (\S\ref{sec:gls-hierarchy}), we discuss how multiple GLS models can be combined to assimilate more of the manual classification tasks.

\subsubsection{SITL-ABS Comparison}
Next, we compare the ABS to all SITL selections (Fig.~\ref{fig:venn-diagram}a) and to only those SITL selections that were classified as MP crossings ((Fig.~\ref{fig:venn-diagram}b). The SITL selects 229 of 278 ABS segments, a larger percentage (82\% vs. 71\%) than GLS segments; however, only 19\% of ABS segments (59 of 278) were classified as MP by the SITL. Conversely, only 28\% of SITL selections and 34\% of MP crossings were selected by the ABS (Fig.~S1e,k). While a majority of ABS selections are of interest to the SITL, the ABS is significantly under-selecting both in general and with respect to MP crossings.

The differences between the ABS and GLS are most likely due to how they were trained. The ABS was trained to select MP crossings that contain EDRs, which typically exhibit larger amplitude variations than MP crossings that do not. Non-EDR time intervals that exhibit large amplitude variations are still of interest to the SITL, but most MP crossings do not exhibit such activity (e.g. the MP crossings in Orbit 1058 shown in Fig.~\ref{fig:orbit-performance}), meaning they are not selected by the ABS.

\subsubsection{ABS-GLS Comparison}
We have been incorporating the ABS into the discussion so far because it is representative of how most other missions with burst memory management systems select data. Here, we compare it to the new GLS machine learning model. While the two systems were trained differently and for slightly different purposes, comparing them may provide a general impression of a) the efficacy of a linear combination of summary data (TDNs) vs. a non-linear combination of a more robust dataset (survey data), and b) models trained for a specific task vs. a potential catch-all model. A comparison may also be influential to future mission designs.

The Venn Diagram shows that only a small fraction of GLS segments were selected by the ABS (83 of 360, or 23\%). Of those, nearly all (81) were selected by the SITL, but only about half (43) were classified as MP crossings. The GLS selects a similarly low fraction of ABS segments (30\%), but of the 53 ABS segments that were also selected by the SITL as MP crossings, the GLS selected 42 of them (Fig~S1f,l). So although both the GLS and ABS under-select compared to all SITL selections, they are not redundant; they each make useful, complementary selections that are highly relevant to the SITL.

\subsection{Performance Metrics}
\label{sec:model-scores}

\subsubsection{Precision, Recall, and F1 Score}
\label{sec:precision-recall-f1}

\begin{figure}[t]
    \centering
    \includegraphics[width=\textwidth]{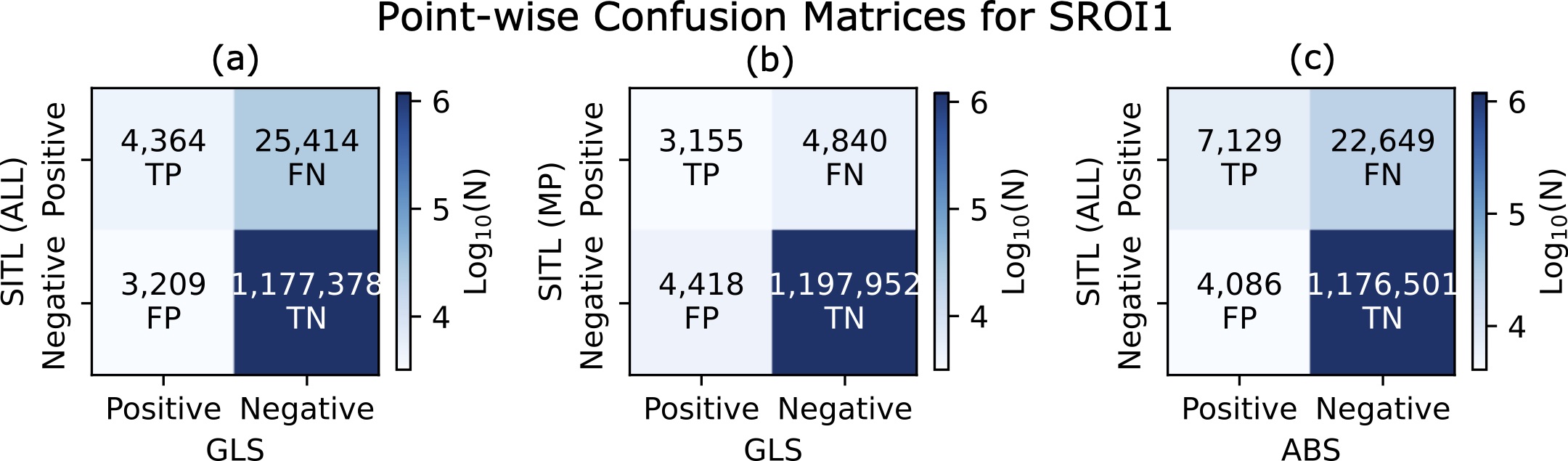}
    \caption{GLS and ABS classifications compared to those of the SITL on a point-by-point basis. Such a metric is typical for machine learning models but does not properly weight predictions with partial or multiple overlap, as when the SITL selects additional context around a given event, or when multiple events are combined into one SITL selection.}
    \label{fig:confusion-matrix}
\end{figure}

Performance of the MP model is directly tied to the threshold filter described in \S\ref{sec:mp-model}. The filter turns the prediction into a binary classifier, where 1 indicates that the observations are from the MP and 0 indicates otherwise. If the model predicts an MP crossing (or not) that was actually classified as a MP crossing by the SITL, this is known as a true positive (false negative). Conversely, if the model classifies the observations as MP (or not) and the SITL does not, this is known as false positive (true negative). Such labels were determined for predicted GLS segments using all SITL selections and only those selections classified as MP crossings, as well as for ABS segments. They are  shown in the form of a confusion matrix in Figure~\ref{fig:confusion-matrix} for SROI1 during the same date range as covered by Figure~\ref{fig:venn-diagram}.

Results from the confusion matrix can be summarized by the precision, recall, and F1 scores. Precision is the fraction of all positive predictions that were correctly classified, while recall is the fraction of all actual positive cases that were correctly predicted and gives a sense of the number of cases missed by the model. The F1 score is a measure that captures the properties of both precision and recall.

The precision, recall, and F1 score for the GLS were (0.58, 0.15, 0.23) for all SITL selections and (0.42, 0.39, 0.41) for just the MP SITL selections. The GLS precision is higher and its recall is lower when compared to all SITL selections, alluding to the MP-like selections that were not classified as MP encounters by the SITL, as mentioned in relation to Figure~\ref{fig:orbit-performance}. From recall, we infer that large fraction of MP points are left unselected by the GLS. This is due, in part, to the fact that the SITL is selecting contextual information tha the GLS is not (again, see Fig.~\ref{fig:orbit-performance}). The F1 score is higher when only MP selections are considered, reflecting the better match between the model and the data considered. 

For the ABS, the precision, recall, and F1 score were (0.64, 0.24, 0.35). High precision and low recall indicates that most ABS selections are important to the SITL, but that the SITL is selecting much more than the ABS. Similar conclusions were deduced from the Venn diagram in Figure~\ref{fig:venn-diagram}. The F1 score is better than that of the GLS when all selections are considered, but lower than when the GLS is compared to only MP points.

\subsubsection{ROC Curve}
\label{sec:roc}

\begin{figure}[t]
    \centering
    \includegraphics[width=0.5\textwidth]{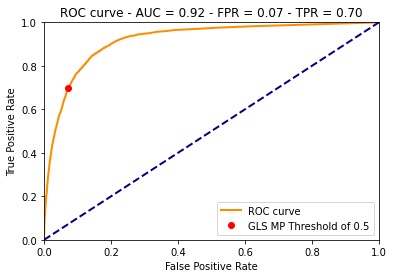}
    \caption{Trade-off between over-selecting false events and under-selecting true events. For MMS, electron diffusion regions are rare and difficult to impossible to observe with the SITL data, so it is willing to over-select by choosing all MP crossings, but it still has to contend with telemetry restrictions. The choice of thresholds for the GLS MP model tries to emulate this approach.}
    \label{fig:model-roc}
\end{figure}

The ability of a model to distinguish between positive and negative cases is indicated by the Receiver Operating Characteristic (ROC) curve, which plots recall (the true positive rate) against the false positive rate, or the number of false positives out of all actual negative cases, for a variety of threshold values. If the area under the ROC curve is 1.0, the positive and negative cases are perfectly distinguishable by the model. If it is 0.5, the model has no ability to distinguish between positive and negative cases. The ROC curve from the validation data of the GLS MP model is shown in Figure~\ref{fig:model-roc}. It has an area under the curve of 0.92, indicating it can tell MP data from non-MP data accurately.

Implementing the model requires a choice in threshold values that involves a trade-off between true positive and false negative rates. The MP model uses a threshold value of 0.5, resulting in a true positive rate of 0.7 and a false positive rate of 0.07. Comparing to the confusion matrix results in Figures~\ref{fig:confusion-matrix}a,b, both the model recall and false positive rates are lower than for the validation set. The model selects fewer SITL-classified points (MP or otherwise), but also makes significantly fewer false predictions than for the validation set.

\section{Discussion}
\label{sec:discussion}
The GLS MP model presented above is the first ML model implemented into the MMS burst management system to automate critical mission operation tasks. To fully automate the burst selection process, the GLS and ABS systems need to be expanded to:

\begin{enumerate}
    \item Identify the variety of phenomena listed in the Seasonal SITL Guidelines
    \item Assign appropriate FOM values to each phenomena
    \item Include an appropriate amount of context around each selection
    \item Respond to external control factors
\end{enumerate}

To classify all phenomena within the SITL Guidelines (Item 1), models could be trained on all SITL selections. However, as the SITL rotates and mission priorities change, model performance would suffer. A better approach would be to create a hierarchy of classification models, as described in \S\ref{sec:gls-hierarchy}.

To assign appropriate FOM values (Item 2), two basic approaches could be considered: use a regression instead of a classification model, or create another model that classifies only on sub-types of MP crossings. In terms of the LSTM MP model, radial basis functions could be used instead of sigmoid functions for activation. Unlike sigmoid functions, radial basis functions map inputs to a continuous output variable so the model could be trained to predict FOM values. These models, however, would have to be retrained whenever mission priorities changed. As an example from MMS, during Phase 3B, low-shear MP crossings were classified as Category 3 events, whereas in all other phases they have been Category 1. To be more adaptable to changing mission priorities, events classified as MP crossings by the MP model could go through another stage of classification that identifies their sub-type (complete/partial, high-/low- shear, etc). The sub-type, then, passes through a look-up table to assign a more appropriate FOM value.

Adding contextual information (Item 3) is relevant to the selections made on orbits 1055 and 1058 SROI1 in Figure~\ref{fig:orbit-performance}. Model predictions could go through some post-processing to simply expand the selection forward and backward in time by a fixed amount or by some percentage of the selection duration.

Outside of the science considerations are operational control factors (Item 4), such as the amount of on-board memory available to store selections. Such considerations often influence the number of selections that the SITL makes. Once GLS selections are made (Items 1 and 3), they can be passed through the GLS Guideline look-up table and assigned an appropriate FOM (Item 2), then filtered through a system monitor that is aware of the state of on-board memory and can make decisions regarding the current set of selections. In some sense, the FOM prioritization does this intrinsically; however, selecting 2.5\,hrs of a low-shear, slow MP crossing could potentially overwrite many other selections.

The GLS is an example of progressive autonomy \citep{Truszkowski:2005}. It follows similar efforts undertaken by NASA to reduce mission costs through greater autonomy in ground control and spacecraft operations \citep{Truszkowski:2006}. Autonomy can alleviate mission complexity and provide real-time decision making when communications latency exists \citep{Truszkowski:2004}. The GLS MP model represents a key advancement toward reducing mission complexity by 1) facilitating larger data rates and more spacecraft through consolidation of event selection processes into a near real-time expandable and adaptable machine learning framework, and 2) accurately identifying and classifying events associated with prime science objectives.


\subsection{Ground Loop Hierarchy}
\label{sec:gls-hierarchy}
\begin{figure}
    \centering
    \includegraphics[width=\linewidth]{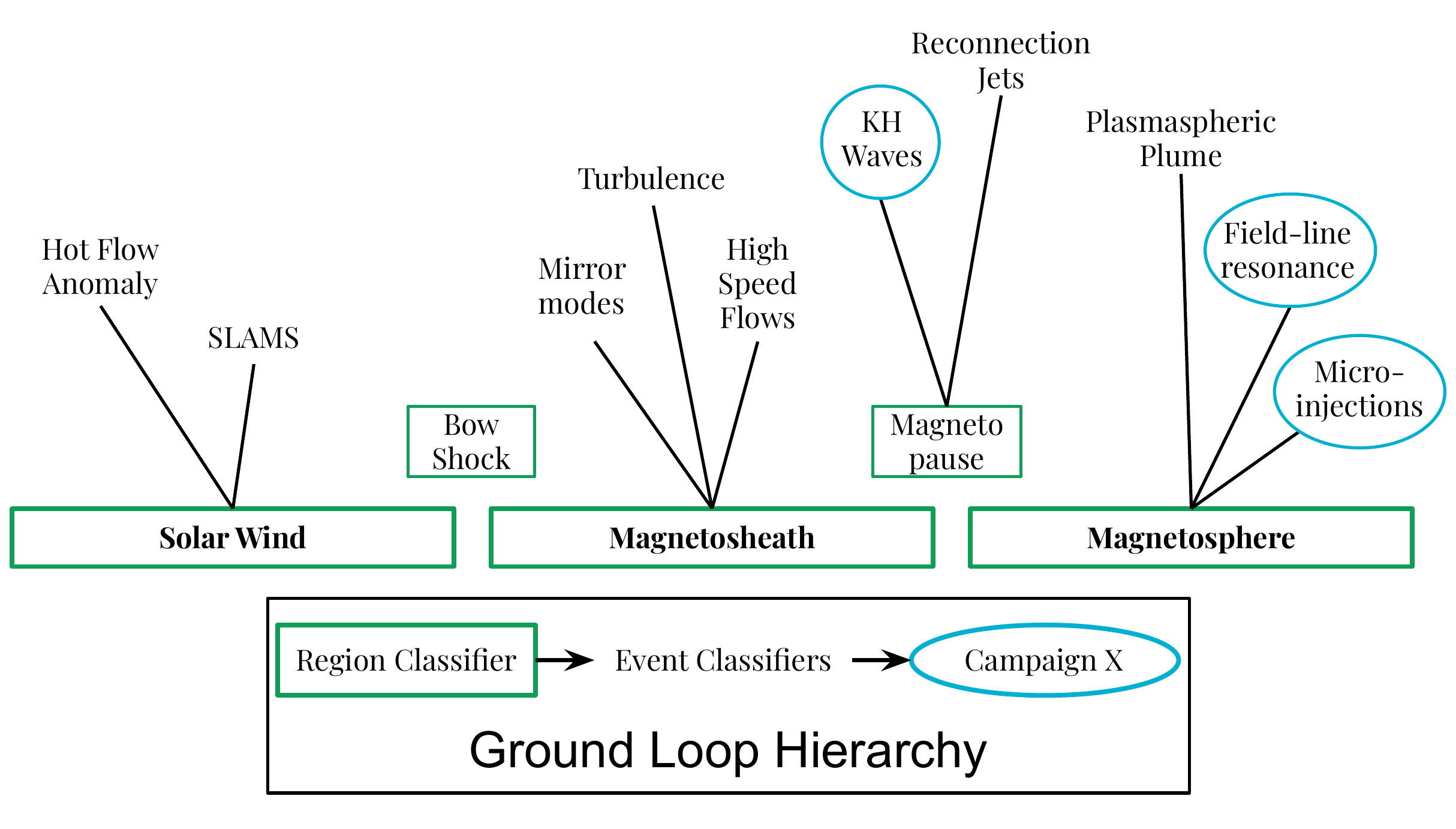}
    \caption{A hierarchy of machine learning models to automate science objectives and reduce mission costs. Data is filtered through region classifiers (green rectangles) and passed to specialized event classifiers (black text) specific to those regions. Multiple event classifiers are activated together to automate science campaigns (blue ovals).}
    \label{fig:ground-loop-hierarchy}
\end{figure}

In the design phase, it was always envisioned that the ABS and GLS would eventually replace the SITL. So now that the first ground loop is in place, what can be done to expand the ground-loop infrastructure for that purpose? 
We propose the Ground Loop Hierarchy.
The Hierarchy follows leaders in industry that found that combining many specialized models often out-performs one comprehensive model (e.g. \citet{Zillow:2015}) For the GLS, this means training a hierarchy of models, as shown in Figure~\ref{fig:ground-loop-hierarchy}. At its lowest level, The Hierarchy consists of region classifiers that segregate data from topologically distinct regions of space. Tier 2 of the hierarchy consists of event classifiers that identify phenomena that are peculiar to a specific region. A third tier could distinguish between similar events to assign more appropriate FOM values, as suggested by the SITL Guidelines for MP crossings (Table~\ref{tab:sitl-guidelines}). The final tier then activates sets of event classifiers from Tier 2 to answer science questions.

Applying the Ground Loop Hierarchy to the dayside region of Earth, all of the SITL data would be passed through the region classifiers to identify which data was recorded in the solar wind, magnetosheath, and magnetosphere, with the bow shock and magnetopause data being classified as the transitions between regions. Next, solar wind data would be passed to the Hot Flow Anomaly (HFAs) event classifier to identify HFAs. A similar process would be applied to all event classifiers, such as to identify mirror mode structures in the magnetosheath, plasmaspheric plumes in the magnetosphere, etc. In this way, the model that classifies HFAs does not have to know anything about the magnetosheath or magnetosphere. Finally, as mission objectives evolve from the primary mission through extended missions, different combinations of event classifiers can be activated to adapt to changing science goals or to strategic science campaigns.

As an example of a science campaign, we build upon a recent MMS discovery of micro-injections at the dusk flank MP in conjunction with ULF wave activity \citep{Fennell:2016}. Simulations proposed that Kelvin-Helmholtz waves (KHWs) on the MP surface were the cause \citep{Kavosi:2018}. To gain more insight into this multi-scale process, one could create event classifiers for micro-injections \citep{Claudepierre:2020}, field line resonances, and KHWs. Formulating a science campaign around micro-injections would entail activating each model. The models not only allow the mission to detect a complex series of events, they can also provide additional insights into the nature of the phenomena. Results from such automated science campaigns can be distributed to the wider scientific community in near real time, increasing the potential scientific impact and return of the data.

Work on the Ground Loop Hierarchy is already underway. Several models that could serve as region classifiers have already been developed \citep{Piatt:2019, Olshevsky:2019, Nguyen:2019b, daSilva:2020}, and one is being adapted for that purpose \citep{Piatt:2019}. The LSTM RNN model described above could serve as either a region or an event classifier. Other event classifiers have been developed using MMS data (e.g., \citet{Claudepierre:2020}), but more are needed. Fortunately, the SITL has done the work to manually classify many events in many years of MMS data, and the tools provided as a product of this endeavor \citep{Argall:2020a, Argall:2020b, Argall:2020c, Small:2020} further reduce the effort required to make additions to the GLS. Soon there should be enough event classifiers to create the first automated science campaigns, thereby establishing the Ground Loop Hierarchy.

\section{Summary}
\label{sec:summary}
MMS is providing key insights into the electron dynamics that catalyze the global flow of energy throughout the magnetosphere. Mission-critical science objectives depend on selecting a subset (${\sim}4\%$) of the high time resolution data that fit into its telemetry budget. A burst management system consisting of the Scientist-in-the-Loop (SITL), Automated Burst System, and Ground Loop System (GLS) ensure that the right ${\sim}4\%$ of data makes it to the ground. This paper documents the tools and infrastructure of the burst management system and demonstrates the performance of the first machine learning (ML) model implemented into the GLS to automate the SITL selection tasks. The GLS model is a Long Short-Term Memory Recurrent Neural Network trained on historical SITL selections to classify the magnetopause (MP), a primary task for the SITL as the MP is a key location for studying electron dynamics associated with magnetic reconnection. Since being implemented into the near real-time data stream, the GLS MP model has selected 78\% of SITL-identified MP crossings in the outbound leg of its orbit, 44\% more than the ABS. This represents the first attempt to introduce ML into critical mission operation tasks. By expanding the GLS into a hierarchy of ML models, MMS progresses toward full autonomy in its burst management system, thereby reducing operations costs and transferring information and resources back to answering fundamental science questions.



\section*{Contribution to the Field}
This article describes the first application of machine learning into the near real-time data flow to select mission-critical burst data. One major limitation that all space missions face is data telemetry restrictions. To combat these, past missions used burst triggers to activate faster data sampling. Such burst triggers and their efficacy are not well documented in the science or instrumentation literature. The Magnetospheric Multiscale (MMS) mission captures burst data continuously whenever it is in its region of interest and has three burst management systems to select segments for downlink: a traditional on-board Automated Burst System (ABS), a human Scientist-in-the-Loop (SITL), and a Ground Loop System (GLS). We describe the GLS and how a Long Short-Term Memory (LSTM) neural network was trained using historical SITL selections to detect magnetopause crossings, a primary task of the SITL. We assess the efficacy of the GLS model by comparing its near real-time selections to those made by the  SITL and ABS. Our work represents a step toward automating the burst memory management system to reduce mission operations costs.

\section*{Conflict of Interest Statement}
The authors declare that the research was conducted in the absence of any commercial or financial relationships that could be construed as a potential conflict of interest.

\section*{Author Contributions}
M.R.A, C.S., S.P, L.B., and M.P. developed machine learning models and analyzed data. J.B., K.K., and K.L. developed the processes at the MMS SDC to generate, store, and distribute MMS data and burst selections. R.E.E., F.D.W. and M.O. wrote the SPEDAS and EVA software, with contributions from many of the other authors. W.R.P. determined the weights and offsets for the ABS. R.B.T., R.E.E., T.P., B.L.G., J.L.B., F.D.W., and M.O. are super-SITLs responsible for the management of the MMS SITL infrastructure. M.R.A, C.S., M.P., M.O., F.D.W., K.K., and W.R.P contributed to the writing of the paper. All authors helped build the ground loop infrastructure.

\section*{Funding}
This work was supported by NASA grant 80NSSC19K1203 and contract NNG04EB99C.

\section*{Acknowledgments}
M.R.A would like to thank M. G. Bobra for inspiration and fruitful discussions, S. C. Zaffke for helpful commentary, and all of the MMS SILTers for their collective work in classifying magnetopause crossings.

\section*{Data Availability Statement}
The datasets analyzed for this study, including the Level 2 data products plotted in Figure~\ref{fig:orbit-performance} and the ABS, GLS, and SITL selections are publicly available through the MMS SDC (\href{https://lasp.colorado.edu/mms/sdc/public/}{https://lasp.colorado.edu/mms/sdc/public/}) and are accessible via PyMMS \citep{Argall:2020a} or the MMS-Plugin for SPEDAS \citep{Angelopoulos:2019}. Notebooks to recreate the tables and figures presented in this manuscript \citep{Argall:2020b}, to train and run the GLS MP model \citep{Small:2020}, and the training and validation data, model weights, and scaling factors \citep{Argall:2020c} are also available through the open access repository Zenodo.

\bibliographystyle{frontiersinHLTH_FPHY} 
\bibliography{frontiers_gls}


\end{document}